# PARTICULATE CONTAMINATION WITHIN FUSION DEVICES AND COMPLEX (DUSTY) PLASMAS

J. Creel[*], J. Carmona-Reyes, J. Kong and
Truell W. Hyde[*]

*CASPER, One Bear Place 97310,
Baylor University, Waco, Texas 76798-7310 USA*

## Abstract

Over the past decade, dust particulate contamination has increasingly become an area of concern within the fusion research community. In a burning plasma machine design like the International Thermonuclear Experimental Reactor (ITER), dust contamination presents problems for diagnostic integration and may contribute to tritium safety issues. Additionally due to ITER design, such dust contamination problems are projected to become of even greater concern due to dust/wall interactions and possible instabilities created within the plasma by such particulates. Since the dynamics of such dust can in general be explained employing a combination of the ion drag, Coulomb force, and ion pre-sheath drifts, recent research in complex (dusty) plasma physics often offers unique insights for this research area. This paper will discuss the possibility of how experimental observations of the dust and plasma parameters within a GEC rf Reference Cell might be employed to diagnose conditions within fusion reactors, hopefully providing insight into possible mechanisms for dust detection and removal.

## I. BACKGROUND

Plasma was once defined as the fourth state of matter and described a fully ionized gas with approximately equal numbers of positive ions and negative electrons. Over time, with the advent of new technology and further experimentation, plasma physics has been redefined as a system of particles whose properties and behavior are dominated by collective long-range interactions [1]. A complex (dusty) plasma is a plasma which also contains charged dust particles ranging in size from nanometers to micrometers [2]. Dust particles can acquire a positive or negative charge due to collisions with the plasma, and tend to be negatively charged due to the increased mobility of electrons as compared to the ions.

The study of complex plasmas has been of increasing importance over recent years with a need for a common experimental apparatus recognized in the late 1980s. As such, a standard platform for studying laboratory dusty plasmas was developed known as the Gaseous Electronics Conference (GEC) rf reference cell [3].

Fusion on the other hand has been around since the beginning of time, or at least since the first star joined the main sequence. Ever since scientists could understand the source of energy that powered the stars, controlled fusion experiments have been an attractive means for clean, cheap energy. However, for a variety of reasons, sustained fusion as a viable energy source has not yet been achieved.

Over the past several decades, a large amount of research has been dedicated to two of the more attractive fusion processes, inertial confinement fusion (ICF) and magnetic confinement fusion (MCF). Within the last decade, particular interest has been focused on the production of dust particulates within these fusion machines. The manner in which this dust affects the physics of either of these systems has yet to be determined.

For this reason, this paper undertakes an initial evaluation of the problem in an attempt to compare and contrast the operational parameters surrounding ICF and MCF with the goal of hopefully identifying a comparable operational range within a GEC rf Reference Cell. If such a range exists, experiments based within a GEC cell might shed light on the specific role that dust particulates play in the fusion process, providing methods for better detection and removal. This information should become be of even greater interest within the next decade as ITER comes on-line.

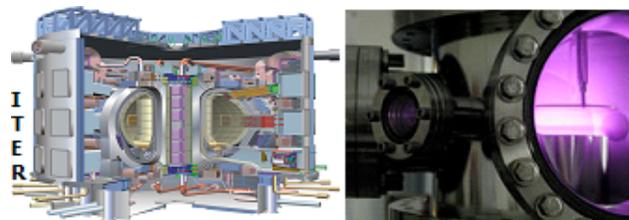

**Figure 1.** The plasma regime near the chamber wall of a fusion device is the region most closely resembling the plasma, dust and boundary conditions within a typical GEC rf Reference Cell. (a) Image of the ITER chamber photo by ITER.org and (b) GEC rf Reference Cell.

---

[*] email: James_Creel@baylor.edu
[*] email: Truell_Hyde@baylor.edu

## II. DISCUSSION

Dust can be created within fusion devices in a variety of ways ranging from production due to high heat fluxes on chamber walls to condensing vaporized material from ejected wall material or initial contamination. Such dust production can create safety concerns due both to the subsequent production of unforeseen operational abnormalities in the plasma as well as the creation of radiological and chemically reactive dust species. As within a typical GEC rf Reference Cell, such dust interacts with the surrounding plasma, acting as 'sources' or 'sinks' to the plasma. As a result, dust influences both the global and local plasma parameters often creating plasma instabilities.

Thus, despite the obvious differences in plasma temperatures existing between complex and fusion plasmas, there are areas within a fusion device that can be quite similar to those within a reference cell. In particular, if the areas around the wall of the fusion device are considered, low plasma temperatures as well as plasma material interactions of the same order as seen in a GEC cell are often prevalent.

## III. DUST DETECTION AND REMOVAL

In addition to the above, diagnostic methods designed to measure the basic operating conditions within a plasma can also be influenced by dust in that in-situ probes can become coated with particulate material and optics can become obscured. This is particularly troublesome for operational procedures relying on optics; for example, such problems can create a non-uniform illumination of the pellet (ICF). Due to the diagnostic problems listed above, observation of the dust is often a problem. The high luminosity of the plasma makes dust (which is often in the submicron size regime) undetectable by conventional methods. Unfortunately, identification of the dust as well as an accurate measurement of its location is absolutely essential for proper analysis or dust removal. One recently developed dust mitigation method involves an electrostatic 'vacuum cleaner' composed of an electrostatic source that is attractive to charged dust particles. Other methods for removal involve Gas Agitation and Vent (GAAV) and Atomic O with RF-assisted discharge (O+RF), both of which are similar to the role of the 'vacuum cleaner' [4].

## IV. DUST PARAMETERS

In order to optimize dust removal employing any of the techniques discussed, the dust particulate parameters (i.e., chemical composition, particulate dynamics, size, charge, etc.) must first be known. Although some of this information can be extracted from in-situ measurements, the majority cannot. This problem has been addressed previously through tactile analysis methods such as SEM, STM, EDAX, or by simply separating magnetic from non-magnetic particles using a magnetic field [7]. Optical methods involving high-resolution microscopy in order to see the structure of the particle itself have also been used. Interestingly, much of the dust found in fusion environments to date is similar in size, shape and composition to the microspheres commonly used within complex plasma experiments employing a GEC reference cell.

The source of the dust found within fusion systems is also of great interest. Plasma arcing to the containment walls ablates wall material, creating changes in the overall wall geometry. This has the ability to appreciably influence global plasma parameters due both to the resulting imperfections in the boundary conditions and overall contamination of the plasma.

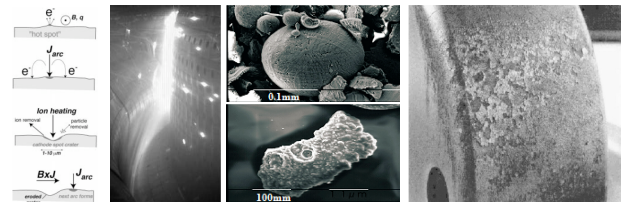

**Figure 3.** (a.) The creation of dust particulates due to plasma arcing [7]. (b.) Particles at high temperature within the bulk plasma [8] (c.) Microscopic images of dust particles within a fusion chamber [9]. (d.) Section of a degraded polodial limiter [10].

## V. SAFETY CONCERNS

When contained within a fusion vessel at low number densities, dust is generally not a safety issue. As a result, most dust related research to date sought primarily to establish limits on dust production as a mechanism for providing 'good housekeeping' within the fusion chamber. More recently however, much larger concerns have been raised concerning dust production within ITER. Higher operating powers coupled with increased wall surface areas implies dust production will be particularly problematic in this environment.

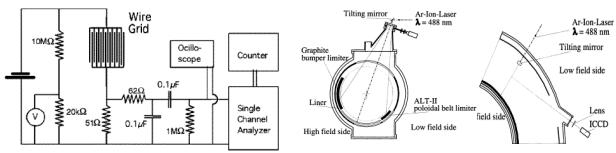

**Figure 2.** Above are two in-situ dust detection methods. (a) A grid which creates a short circuit upon dust impact [5] and (b) an optical detection method relying on laser illumination of dust particles [6].

As mentioned earlier, dust species within the burning plasma of a fusion reactor also has the ability to become chemically or radiologically reactive. This creates additional concerns during chamber venting since explosions can occur from the reaction of such dust with atmospheric hydrogen. Mobilizations of particles during the venting process or from explosions occurring due to vacuum leaks are also areas of concern.

## VI. DUST / PLASMA INTERACTIONS

As mentioned above, it is well know from experiments within the complex plasma community that any dust within a plasma interacts with the surrounding plasma. (In complex plasmas, this reaction is large enough that the dust is considered an integral component of the plasma.) Many of these interactions may have possible applications within fusion systems. For example, recent GEC data suggests that increasing dust production appears to increase conductivity between the plasma and the wall leading to arcing and new dust production. Dust precipitating on surfaces, or crevices created by particle dislodgment can also change the overall engineering design of the chamber impacting both plasma performance and stability. Dust acting as a sink/source within the plasma, also impacts the ion recombination rate and can emit electrons due to photoionization or thermal emission, creating positively charged particulate sources [2].

Dust dynamics within a GEC rf Reference Cell is characterized by a variety of forces [2]. Despite the small size of the particles, gravity plays a critical role in determining the overall dynamics of the dust, establishing the levitation height of the dust. Obviously, the charge on the dust (due primarily to electron bombardment and plasma ion flows) also contributes to the final equilibrium symmetry (if any) of levitated dust. Intrinsic characteristics of the experimental chamber such as the rf driving frequency, variable bias, and the operating neutral gas pressure also create fluctuations in the particle charge ultimately affecting overall dust particle behavior. Depending on the system screening length and potential, all of the above contribute to overall dust behavior.

In fusion environments, additional dust charging mechanisms are present where particle charging by radioactive emission can cause a particle to become positively charged. Experiments relating to this type of charging have been conducted in a GEC rf Reference Cell, another example of potential crossover between the two fields [11]. Finally, particle dynamics within either a reference cell or a fusion device are strongly influenced by thermal forces. The bulk plasma within either system is generally hotter than the walls, facilitating dust and plasma diffusion. Again, these types of dynamics have been investigated at length in GEC rf Reference Cells.

## VII. CONNECTIONS

A large amount of data has been collected on particle/plasma interactions within a typical GEC rf Reference Cell. It would be beneficial to both the fields of fusion and complex (dusty) plasmas to develop a consistent parameter list allowing comparison between the two. As discussed above, the plasma sheath boundary area is common to both and could possibly be scaled accordingly. There are a large number of areas within fusion plasmas where the need for additional information of this type is essential; many are also areas which might lead to a mutually beneficial partnership between complex and fusion plasmas. Several fusion research areas [7], [10] have recently been cited that might well overlap with experiments conducted within a GEC rf Reference Cell. Examples include: determining the temperature stress of different materials and their corresponding reactions; establishing sputtering limits and examining the fractal growth of particles; dust removal methods; modeling of dust transport mechanisms particularly those geared toward the prevention of accidental release of particles; accretion from supersaturated vapor; varying the particle charge through photoemission; the removal of dust from shadowed surfaces within fusion devices; and the understanding of the complex linear feedback mechanisms present between plasma boundaries and materials [7], [10].

## VIII. CONCLUSION

A brief overview of a few of the dust production mechanisms within fusion devices along with an attempt to identify possible overlap areas of interest within complex (dusty) plasmas has been given. It is hoped that an expansion of this overview might lead to research opportunities beneficial to both fields. The most obvious candidate of this sort currently consists of an examination of the physics describing the sheath region within low energy complex plasmas and edge fusion plasmas. However other areas of overlap appear to exist, for example those involving dust contaminant production and its impact on system diagnostic capabilities. As discussed above, it is hoped that such overlap areas between the two fields might lead to a better understanding of the basic physics behind dust production, detection and removal across both regimes.